Subj-class: Material Science

\documentstyle[prl,aps,psfig,floats]{revtex}

\begin{document}
\draft 

\twocolumn[\hsize\textwidth\columnwidth\hsize\csname@twocolumnfalse\endcsname

\title{The structure and phase stability of CO adsorbates on Rh(110)}

\author{Dario Alf\`e$^1$ and Stefano Baroni$^{1,2}$}

\address{$^1$INFM -- Istituto Nazionale di Fisica della Materia and \\
SISSA -- Scuola Internazionale di Studi Superiori ed Avanzati, via
Beirut 2-4, I-34014 Trieste, Italy\\
$^2$CECAM, ENSL, Aile LR5, 46 All\'ee d'Italie,
F-69364 Lyon Cedex 07, France. }

\date{\today}
\maketitle

\begin{abstract} The structure of CO adsorbates on the Rh(110) surface
is studied at full coverage using first-principles techniques. The
relative energies of different adsorbate geometries are determined by
means of accurate structure optimizations. In agreement with
experiments, we find that a $\rm p2mg(2\times 1)$--$\rm 2CO$ structure
is the most stable. The CO molecules sit on the short-bridge site
(carbon below) with the molecular axis slightly tilted off the surface
normal, along the (001) direction. Configurations corresponding to
different distributions of tilt angles are mapped onto an anisotropic
2D Ising model whose parameters are extracted from our {\it ab-initio}
calculations. We find that an order-disorder phase-transition occurs
at a temperature $T_c \approx 350~^{\rm o}{\rm K}$.

\vspace{0.4cm}
\end{abstract}

]

\narrowtext

\section{Introduction} Rhodium surfaces are attracting a wide
scientific and technological interest due to their catalytic
properties, particularly because they act so as to reduce the energy
activation barrier for the reaction $2{\rm CO} + 2{\rm NO} \rightarrow
2{\rm CO}_{2} + {\rm N}_{2}$, and thus to eliminate the two poisonous
CO and NO gases from the pollution emission of combustion engines.

The stable structure of the Rh(110) clean surface is
unreconstructed. However, if prepared in a convenient way with oxygen
adsorption it may also present---upon thermal desorption---meta-stable
$(1\times n),~~( n = 2, 3, 4, 5 )$, missing- or added-row structures
which revert to the unreconstructed one at temperatures above
$480~^{\rm o}{\rm K}$ \cite{CoDhKiPaPaPrRo,DhCoCaPaKiPrRo,AlRuKiRo}. 

The adsorption of CO molecules on Rh(110)
has been studied by means of a variety of techniques
\cite{BoGuJo,ScKr,BaDhCoPrRo,DhBaCoPaKiRo,CaBaCoDhKiRo,BaBaStVHSo,WeLoBl,PrSaMoCo,GoRoPeFeAl}.
The adsorption of $1$ monolayer (ML) of carbon monoxide on the
unreconstructed surface results in a $(2\times 1)$p2mg structure, with
the C atom bound in the short bridge sites along the $(1\bar 10)$
direction, and the molecular axis alternatively tilted with respect to
the surface normal, towards $(001)$ direction \cite{BaBaStVHSo}. In
Ref. \cite{BaDhCoPrRo}, the $(2\times 1)$p2mg LEED pattern was
reported to disappear at temperatures higher than $\approx 270\div
280~^{\rm o}{\rm K}$, well below the desorption of CO from the
surface. This fact was tentatively explained in terms of an
order-disorder phase-transition.

In this paper the structure and phase stability of one ML of CO
molecules adsorbed on the Rh(110) $(1\times 1)$ surface are studied
from first principles and by mapping the low-lying energy
configurations corresponding to the different distributions of tilt
angles onto an anisotropic 2D Ising model. The latter is then
simulated using a standard Metropolis Monte Carlo algorithm. The
order-disorder transition temperature estimated from our simulations
is $\approx 350\pm 50~^{\rm o}{\rm K}$ , in fair agreement with
experimental findings \cite{BaDhCoPrRo}.

\section {Computational method} Our calculations are based on density
functional theory within the local-density approximation (LDA)
\cite{HoKo64,KoSh65}, using Ceperley-Alder correlation energies
\cite{CeAl}. The one-particle Kohn-Sham equations are solved
self-consistently using plane-wave (PW) basis sets in a
pseudopotential scheme. Because of the well known {\it hardness} of
the norm-conserving (NC) pseudopotentials for the O and---to a lesser
extent---Rh atoms, we make use of {\it ultra-soft} (US)
pseudopotentials \cite{Va90} which allow an accurate description of
the O and Rh valence pseudo-wavefunctions with a modest basis set
including PW's up to a kinetic-energy cutoff of 30 Ry. In the case of
Rh, we found it convenient to treat the $s$ and $p$ channels using a
NC potential, while the US scheme is applied only to the {\it hard}
$d$ orbital \cite{stokbro,StBa}. With such a small basis set, the
accuracy is slightly improved if also C is treated within the US
scheme, which we decided to do. Brillouin-zone (BZ) integrations are
performed using the Gaussian-smearing \cite{MePa} special-point
\cite{MoPa} technique. In agreement with Ref. \cite{StBa}, we find
that the structural properties of bulk rhodium are well converged
using a first-order Gaussian smearing function \cite{MePa} of width
$\sigma=0.03 \rm ~Ry$ and $10$ special {\bf k}-points in the
irreducible wedge of the BZ (IBZ). The isolated surface is modeled by
a periodically repeated supercell. We have used the same supercell for
both the clean and the CO-covered surfaces. For the clean surface we
have used $7$ atomic layers plus a vacuum region corresponding to
$\approx 9$ layers. For the CO-covered surface the $7$ Rh layers are
completed by one layer of CO molecules on each side of the slab: in
this case the vacuum region is correspondingly reduced to $\approx
5.5$ atomic layers. We have used the same Gaussian-smearing function
as in the bulk calculations with $8$ special {\bf k}-points in the
surface IBZ. Convergence tests performed with a value of $\sigma$
twice as small and a correspondingly finer mesh of special points
resulted in no significant changes in total energies and equilibrium
geometries. The latter are found by allowing all the atoms in the slab
to relax until the forces acting on each of them are smaller than
$0.5\times 10^{-3} {\rm Ry}/{\rm a_{0}} $.

\section {Results} \subsection{Structural analysis} The clean Rh(110)
surface is unreconstructed. An analysis of LEED data suggests that the
top interlayer spacing is reduced by $6.9\pm 1.0\%$ relative to the
bulk interlayer spacing, while the second interlayer spacing would
expand by $1.9\pm 1.0\%$ \cite{NiBiHaFrSo}. Our {\it ab initio} data
indicate a relaxation of $-9\%$ and of $3.5\%$ in the first and second
interlayer spacings respectively.

\begin{figure}
\centerline{\hbox{\psfig{figure=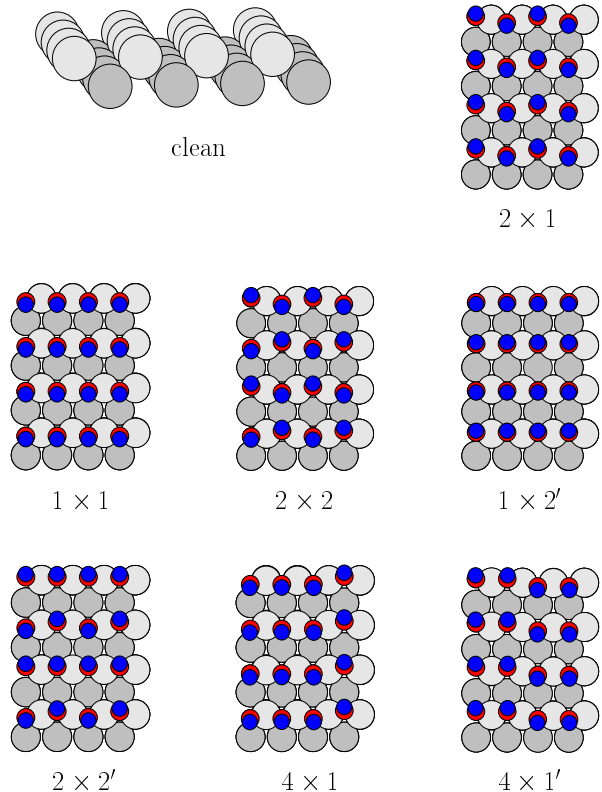,height=4.2in}} }
\caption{ The seven different surface structures referred
in the text and the clean Rh(110) surface. \label{fig:prima}}
\end{figure}

For the CO-covered surface, LEED data indicate that the molecules are
bound in the short bridge site between two first layer rhodium atoms
in the $(001)$ direction with the molecular axis tilted by $24$
degrees from the surface normal, forming a $(2\times 1)\rm p2mg$
structure. In principles there are many different ways to arrange the
CO molecules so as to obtain the same LEED pattern. We concentrate our
attention on three possible adsorption sites: {\it i)} the {\it
short-bridge} one described above; {\it ii)} the {\it on-top} one, in
which the CO molecule is located on top the first-layer atoms; and
{\it iii)} the {\it hollow} site, formed by two first-layer atoms in
the $(001)$ direction and one second-layer atom. In agreement with the
outcome of the LEED analysis, we find that the short-bridge site is
the most favorable. The relative energies of the other two sites with
respect to the short bridge---assuming a $1\times 1)$ structure in all
cases---are: 0.19~eV (hollow) and 0.34~eV (on top). We find that the
angle between the surface normal and the CO molecular axis is $\alpha
= 17\pm 2 $ degrees, and that the angle between the Rh-C bond and the
surface normal is $\delta = 13\pm 2 $ degrees; the Rh-C bond length is
$2.02 \rm~\AA$ and the C-O distance is $1.17\rm~\AA$. The rhodium
substrate presents an outward relaxation of the first layer of $2.8
\%$ with respect to the bulk interlayer spacing. These results are
summarized in Table~\ref{tab:prima} together with similar ones
obtained for six other different surface geometries (See
Fig.~\ref{fig:prima}).

From Table~\ref{tab:prima} we see that the $(1\times 1)$ and $(1\times 2)$
geometries are degenerate within our error bar which we estimate to be
$\pm 1~\rm meV$, and that the uncertainty about the corresponding tilt
angle is very large. This behavior can be understood by a simple
qualitative model of the surface energetics which also accounts for
the observed ordering of the structures. In order to disentangle the
relative importance of the adsorbate-substrate and adsorbate-adsorbate
interactions, we have modeled the former by a $(2\times 2)$ supercell in
which a single CO molecule is constrained to sit at the same
short-bridge site which would be preferred at full coverage.

\begin{table}
\caption{Structural data for seven different
surface structures (see Fig. 1). $\alpha$ and $\delta$ are
respectively the angles between the surface normal and the C-O and the
Rh-C axis, $\Delta E$ is the energy difference per molecule between
the $(n\times m)$ and the $(2 \times 1)$ structures.
The theoretical error is estimated to be $\approx 2 ~\rm meV$.
The experimental values refer to the $(2 \times 1)$
structure} \label{tab:prima}
\begin{tabular}{lccccc}
 & $\alpha$ & $\delta$ & $d(C-O)$ & $d(Rh-C)$ & $\Delta E$ \\
unit & degrees & degrees & $\rm\AA$ & $\rm\AA$ & meV/mol \\
\tableline
Expt. & $24\pm 4\tablenotemark[1]$ & $13\pm 4\tablenotemark[1]$ & 
$1.13\pm 0.09\tablenotemark[1]$ & $1.97\pm 0.09\tablenotemark[1]$ & \\
$2 \times 1$ & $17 \pm 2$ & $13 \pm 2$ & $1.17$ & $2.02$ & $ 0.0 $ \\
$1 \times 1$ & $\lesssim 10$ & $\lesssim 10$ & $1.17$ & $2.02$ & $ 33.5 $ \\
$1 \times 2$ & $\lesssim5$ & $\lesssim 5$ & $1.17$ & $2.02$ & $ 33.5 $ \\
$2 \times 2$ & $13 \pm 2$ & $11 \pm 2$ & $1.17$ & $2.02$ & $ 13.5 $ \\
$2 \times 2^{\prime}$ & $13 \pm 2$ & $11 \pm 2$ & $1.17$ & $2.02$ & $ 21.5 $ \\
$4 \times 1$ & $16 \pm 2$ & $13 \pm 2$ & $1.17$ & $2.02$ & $ 17.0 $ \\
$4 \times 1^{\prime}$ & $16 \pm 2$ & $12 \pm 2$ & $1.17$ & $2.02$ & $ 17.0 $ \\
\end{tabular}
\tablenotetext[1]{From Ref. \cite{BaBaStVHSo}}
\end{table}

We observe that the dependence of
the adsorption energy on $\alpha$ is very weak up to $\alpha\approx
10^{\rm o}$, and that it becomes very steep above this angle. When the
coverage increases, the dipole-dipole interaction becomes important and
accounts qualitatively for the energy ordering of the structures
displayed in Fig.~1. In the $(2\times 1)$ structure, nearest-neighbor
molecules are tilted by opposite angles around the axis joining them
(the $(1\bar 10)$ direction), while next-nearest-neighbor molecules
are tilted by a same angle about the (001) axis which joins them. The
dipole-dipole interaction favors both these arrangements of
angles. The $(2\times 2)$ geometry is similar to the previous one as
regards the nearest-neighbor interactions, whereas it is unfavored
regarding next-nearest-neighbor interactions. The next higher energies
are those of the $(4\times 1)$ and $(4\times 1')$ structures which are
almost degenerate because they have the same number of unlike tilt
angles along the $(1\bar 10)$ row. The next structure is the $(2\times
2')$ one which is characterized by an alternating arrangements of
energetically favored and disfavored rows and columns of CO
molecules. Finally, in the $(1\times 1)$ and $(1\times 2)$ structures the
nearest-neighbor molecules are tilted by a same angle and the
corresponding dipolar interaction is therefore independent of
$\alpha$; it is only the weaker next-nearest-neighbor interaction
which depends on $\alpha$, more so for the $(1\times 2)$ structure for
which the sign of the dipole-dipole interaction energy is the same as
that of the adsorbate-substrate interaction, while the two
interactions tend to cancel for the other structure. In both cases,
this behavior results in a very weak dependence of the energy upon
$\alpha$, and in an energy degeneracy of the two structures, within
our error bars. We have also calculated the adsorption energy of the
CO molecules defined as $E_{slab}^{\rm Rh-CO} - E_{slab}^{\rm Rh} -
E_{slab}^{\rm CO}$, where $E_{slab}^{\rm Rh-CO}$ is the total energy
of the CO covered surface, $E_{slab}^{\rm Rh}$ is the total energy of
the clean Rh(110) surface, and $E_{slab}^{\rm Rh-CO}$ is the total
energy of the CO, all the calculations being done using the same slab
geometry and the same set of {\bf k}-points. The calculated adsorption
energy is of $2.78$ eV/molecule, which has to be compared to the
experimental value $1.1$ eV/molecule \cite{BoGuJo}. This large
discrepancy is a common feature of the LDA which is well known to
overestimate absolute binding energies, whereas equilibrium geometries
and energy differences among them are usually predicted with a much
higher accuracy (of the order of a few percents).

\subsection{Finite-temperature properties} \subsubsection{Mapping onto
a 2D Ising model} From Table~\ref{tab:prima} we see that the energy
necessary to tilt the angle of a molecule is of the order of
$10\div 30~{\rm meV}$, whereas the energy difference between different
adsorption sites is typically ten times as large. This fact indicates
that---given the adsorption sites of the CO molecules---a well defined
energy would correspond to each distribution of tilt angles, where every
molecule is labeled by a variable which indicates in which direction
it is tilted (the sign of the tilt angle, which we associate to an
Ising-like variable, $\sigma=\pm 1$). Much in the same spirit of
the {\it cluster expansion} of the energy landscape of an alloy
\cite{CW}, the energy of each tilt-angle configuration can be
expressed in terms of polynomials in the $\sigma$'s. Because, of
symmetry, odd-power polynomials are absent from the cluster
expansion. Restricting ourselves to second-order polynomials (spin-pair
interactions) and neglecting all the couplings beyond
next-nearest-neighbors, the cluster expansion of the surface energy
would read: $$\displaylines{\quad E[\{\sigma\}] = \frac{1}{2}
\sum_{i,j} \sigma_{i,j} \left ( J_{x}\sum_{\delta = \pm 1}
\sigma_{i+\delta,j} \right . \hfill\cr \hfill \left . +
J_{y}\sum_{\delta =\pm 1} \sigma_{1, j+\delta}
+ \frac{1}{2}J_2
\sum_{\delta,\delta'=\pm 1} \sigma_{i+\delta,j+\delta'} \right ).
\quad (1) }$$ It
is straightforward to see that:
$$\vbox{
\halign {\tabskip=0pt $ # \hfill $ & $\hfill # \hfill $
& $# \hfill $ & \quad $# \hfill $ & $\hfill # \hfill $ & $# \hfill $
\tabskip=0pt plus1fil & \hss\llap{#}\tabskip=0pt \cr E_{2 \times 1} & = &
J_{y} + J_{x} - J_{2}; & E_{1 \times 1} & = & J_{x} + J_{y} +
J_{2};&\cr E_{1 \times 2} & = & J_{x} - J_{y} - J_{2}; & E_{2 \times
2} & = & J_{2} - J_{y} - J_{x}; &
\cr E_{4 \times 1} & = & E_{4
\times 1^\prime} = J_{y}; & E_{2 \times 2^\prime} & = & 0,&\cr }
}\eqno(2)$$
where the subscripts refer to the structures of Fig.~\ref{fig:prima}.

$E_{2 \times 1}$ is the ground-state energy which we take as the
reference energy. The $(4 \times 1)$ and $(4 \times 1^{\prime})$
structures are degenerate within the present model, and their energy
difference provides therefore an estimate of longer-range or many-spin
interactions which have been neglected. Out Eqs. (2), one can extract
four independent energy differences, which are linear functions of the
three parameters $J_x$, $J_y$, and $J_2$. By disregarding one of these
equations in turn, one obtains 4 different linear sistems for the
$J$'s which provide estimates for these parameters, which coincide
within $\approx 1.5~\rm meV$. The average of the four set of
parameters so obtained is: $J_{x} = 13.6 ~\rm meV$, $J_{y} = -3.4 ~\rm
meV$, and $J_{2} = 3.4 ~\rm meV$. Note the large difference between
the absolute values of $J_{x}$ and $J_{y}$, which is due to a stronger
coupling in the `ziz-zag' $(\bar 110)$ direction, where the distance
between neighboring molecules is smaller by a factor $\sqrt{2}$ than
in the orthogonal direction.

\subsubsection {Monte Carlo simulations} 
\begin{figure}
\centerline{\hbox{\psfig{figure=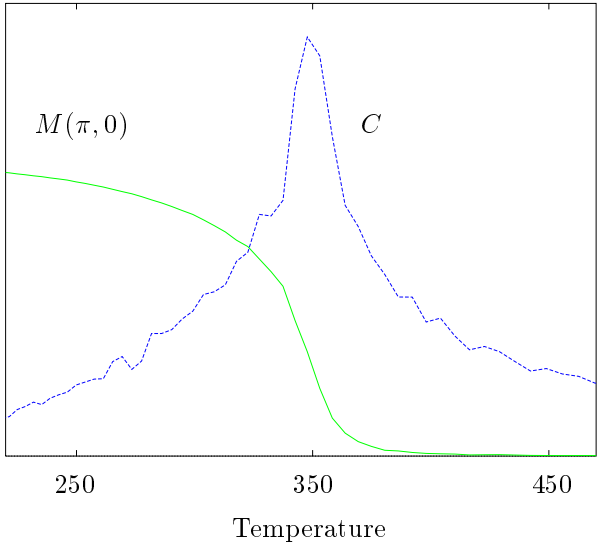,height=3in}} }
\caption{ Specific heat and Fourier transform of the correlation length 
for the 2D spin model (see text). \label{fig:seconda}}
\end{figure}

The thermal properties of our
system are obtained by standard Metropolis Monte Carlo simulations of
the above Ising model (see for example Ref. \cite{chan}). To this end,
we have used a $40\times 40$ square lattice with periodic boundary
conditions. The order parameter of the transition between the $(2\times
1)$ ordered phase and the disorderd phase where the tilt angles are
distributed at random, is the Fourier coefficient of the spin-spin
correlation function, $ M({\bf q}) = \frac{1}{N} \sum_{\bf r}
e^{i{\bf q} \cdot {\bf r}} \langle \sigma_{\bf r} \sigma_{\bf 0}
\rangle $, at wavevector ${\bf q} = (\pi,0)$. The order-disorder
transition temperature, $T_{c}$, is estimated looking at the maximum of
the specific heat $C$. We have not attempted any finite-size scaling,
but we have verified that the location of the transition temperature
is rather insensitive to the choice of the size of the system, by
making a few simulations for a $64\times 64$ spin matrix.  In
Fig.~\protect\ref{fig:seconda} we show the behavior of the specific
heat, $C$, and the order parameter, $M(\pi,0)$, as functions of
temperature.  We estimate the critical temperature to be $T_{c}
\approx 350\pm 50~^{ \rm o}{\rm K}$. The estimate of the error bar is
based on the uncertainties on the coupling coefficients of the Ising
model.


\end{document}